\documentclass{aa}
\usepackage{txfonts}
\usepackage{graphicx}
\usepackage{natbib}
\bibpunct{(}{)}{;}{a}{}{,}

\def\xmm{{\it XMM-Newton}}
\def\intgr{{\it INTEGRAL}}

\begin{document}
\title{South-West extension of the hard X-ray emission from the Coma cluster}
\author{D. Eckert\inst{1,2}, A. Neronov\inst{1,2}, T. J.-L. Courvoisier\inst{1,2} \& N. Produit\inst{1,2}}
\offprints{Dominique Eckert, \email{Dominique.Eckert@obs.unige.ch}}

\institute{INTEGRAL Science Data Centre, 16, ch. d'Ecogia, CH-1290 Versoix, Switzerland
\and Geneva Observatory, University of Geneva, CH-1290 Sauverny, Switzerland}
\date{Received 09-02-07/ Accepted 14-05-07}

\abstract{}{We explore the morphology of hard (18-30~keV) X-ray emission from the Coma cluster of galaxies.} {We analyse a deep (1.1~Ms) observation of the Coma cluster with the ISGRI imager on board the \emph{INTEGRAL} satellite.}{We show that the source extension  in the North-East to South-West (SW) direction ($\sim 17'$)  significantly exceeds the
size of the point spread function of ISGRI, and that the centroid of the image of the source
in the 18-30~keV band is displaced in the SW direction compared to the
centroid in the 1-10~keV band. To test the nature of the SW extension we
fit the data assuming different models of source morphology. The best fit is
achieved with a diffuse source of elliptical shape, although an acceptable fit can be achieved assuming an additional point source SW of the cluster core. In the case of an elliptical source, the direction  of extension of the source coincides with the direction
toward the subcluster falling onto the Coma cluster. If the SW excess is due to the presence of a point source with a hard spectrum, we show that there is no obvious X-ray counterpart for this additional source, and that the closest X-ray source is the quasar EXO 1256+281, which is located $6.1'$ from the centroid of the excess.}{The observed morphology of the hard X-ray emission  clarifies the nature of the hard X-ray ``excess'' emission from the Coma cluster, which is due to the presence of an extended hard X-ray source SW of the cluster core.}

\authorrunning{Eckert D. et al.}
\titlerunning{South-West extension of the hard X-ray emission from the Coma cluster}

\maketitle

\section{Introduction}
Clusters of galaxies are the biggest bound structures of the universe, and, according to the hierarchical scenario of structure formation, the latest to form. They are filled by a hot ($10^8-10^9$ K) plasma, called Intra Cluster Medium (ICM), and thus radiate in soft X-ray bands through thermal Bremsstrahlung.

Since clusters of galaxies are the latest and biggest structures to form, we expect some of them to be still forming, and experiencing major merging events with smaller clusters. This is the case of the Coma cluster, that is currently merging with the NGC 4839 group. In such events, the merging of the ICM of the two clusters creates shock fronts in which theory predicts that a large population of particles would be accelerated to high energies \citep{sarazin}. This phenomenon should then produce a reheating of the gas and create a higher temperature plasma that would radiate more strongly in hard X-rays. Alternatively, interaction of the population of mildly relativistic electrons that produce the halos of galaxy clusters via synchrotron radiation \citep{feretti} with the Cosmic Microwave Background would then produce hard X-ray emission through inverse Compton processes, and thus add a power-law tail to the spectrum in the hard X-ray domain. Another possible model involves a population of multi-TeV electrons that would radiate in hard X-rays through synchrotron emission \citep{timokhin}. Detection of this hard X-ray excess would help in learning more about the cosmic ray population detected by radio observations. Furthermore, characterization of the morphology of the hard X-ray emission would bring a possible identification of acceleration sites, and since clusters of galaxies are one of the few possible candidates for acceleration of cosmic rays at high energies, it would bring important information on the origin of cosmic rays.

Recent reports of detection of a hard X-ray excess by \emph{Beppo-SAX} \citep{fusco} and \emph{RXTE} \citep{rephaeli} in the Coma cluster appear to confirm the existence of a high energy tail of the spectrum of merging clusters, and thus prove the existence of particle acceleration sites in these clusters. However, these detections are quite weak and controversial \citep{rossetti}, and since the hard X-ray instruments on both \emph{Beppo-SAX} and \emph{RXTE} are non-imaging, contamination by very hard point sources inside the cluster cannot be excluded (e.g. by the central galaxy NGC 4874, NGC 4889 or the QSO EXO 1256+281). In addition, no information on the morphology of the hard X-ray emission was obtained. \citet{ren2} presented an analysis of a first 500 ks set of \emph{INTEGRAL} data and were not able to confirm the presence of a hard X-ray excess or to sensibly constrain the hard X-ray morphology of the source.

In this paper, we use the imaging capabilities of the IBIS/ISGRI instrument \citep{lebrun} to extract information on the hard X-ray emission of the Coma cluster. In Sect. \ref{secima}, we present the results of our imaging analysis of the ISGRI data, and compare them with existing \emph{XMM-Newton} data in the soft X-ray domain. In Sect. \ref{secpif}, we describe a method to analyse extended sources with a coded mask instrument to extract quantitative flux measurements, and apply it to the case of the Coma cluster. In Sect. \ref{secspec}, we present a combined \emph{XMM}/\emph{INTEGRAL} spectrum of the cluster. Finally, discussion of our results is presented in Sect. \ref{secdisc}.

\section{Data}
\label{secdata}

Our analysis covered 401 Science Windows (ScWs) of public data, for a total of 1.1 Ms of observation. We analysed the data with the latest release of the Offline Scientific Analysis (OSA), version 6.0, and eliminated ScWs with a high background level. We used the remaining data to create a mosaic image in the standard way. Table \ref{tab} gives the log of the observation.

\begin{table}[hbt]
\begin{tabular}{|c|c|c|c|}
\hline
INTEGRAL & Observation & No. of & Observing \\
revolution number & dates & pointings & time [ks]\\
\hline
\hline
0036 & Jan 29-31, 2003 & 63 & 140.1\\
0071-72 & May 14-18, 2003 & 135 & 304.5\\
0274-75 & Jan 10-15, 2005 & 57 & 202.4\\
0317-18 & May 19-25, 2005 & 99 & 333.4\\
0324-25 & Jun 9-11, 2005 & 47 & 164.5\\
\hline
\cline{4-4}
\multicolumn{3}{r}{ } \vline& 1,144.9\\
\cline{4-4}
\end{tabular}
\caption{INTEGRAL observation log on the Coma cluster}
\label{tab}
\end{table}

The \xmm\ image is produced using the data of the PN camera taken in June 2000 during the Coma mosaic observation \citep{briel}. We used the SAS software version 6.5 and the background substraction method from the Birmingham group \citep{read} to analyse the data.

\section{Imaging analysis in the hard X-ray domain}
\label{secima}
\subsection{Mosaic image of the Coma cluster}
\label{secmos}

\begin{figure}
\resizebox{\hsize}{!}{\includegraphics{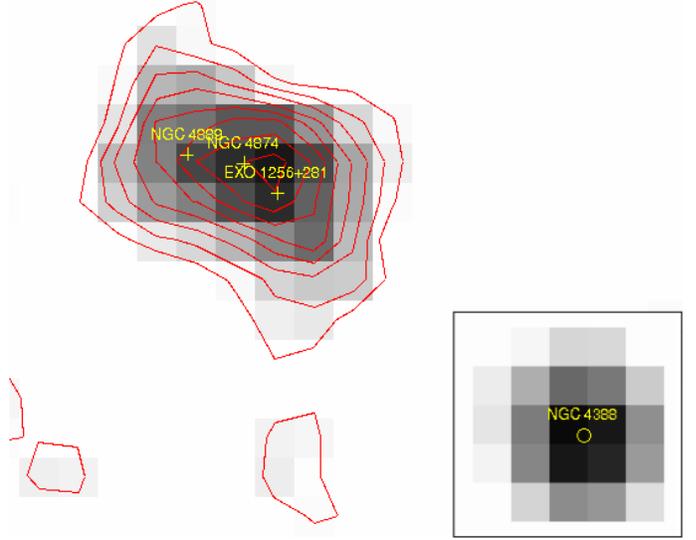}}
\caption{Standard OSA 6.0 significance image of the Coma cluster with $\sim$1.1 Ms of data in the 18-30 keV energy band. Significance contours from 3 to 10$\sigma$ in steps of 1$\sigma$ are overlayed in red. The position of the 3 brightest X-ray point sources is shown. For comparison, the inset in the bottom right corner shows a mosaic image of a known point source in the same field, NGC 4388.}
\label{figosa}
\end{figure}

Figure \ref{figosa} shows a mosaic image in the 18-30 keV band extracted with the standard OSA 6.0 tools, using the data described in Table \ref{tab}, with 3-10 significance contours. For comparison, we show in the inset in the right bottom corner of the image a mosaic image of a point source (NGC 4388) produced using a comparable amount of data and normalized so
that the amplitude of the brightest pixel is the same as in the case of Coma
cluster.

A first look at the image indicates that the Coma cluster source is extended.
It is not surprizing, since the angular size of the core of the of Coma
cluster in 1-10~keV energy band is $D\simeq 20'$ \citep{schuecker}, which is larger than the size of the point spread function (PSF) of ISGRI ($2\sigma=12'$, which is the angle subtended by a 11.2 mm mask hole at a distance of 3200 mm).

\begin{figure*}
\centering
\hbox{
   \includegraphics[width=9cm]{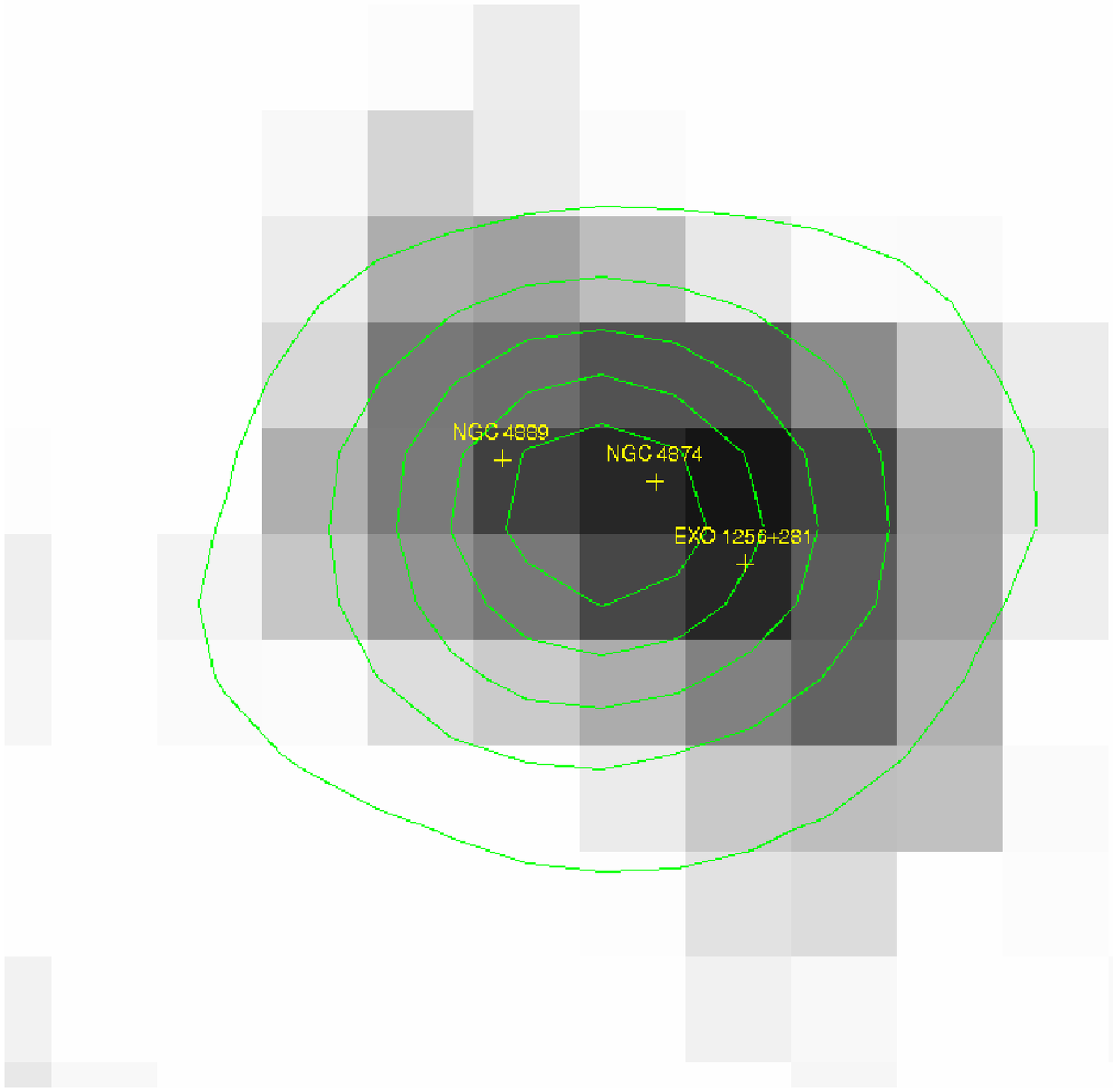}
   \includegraphics[width=9cm]{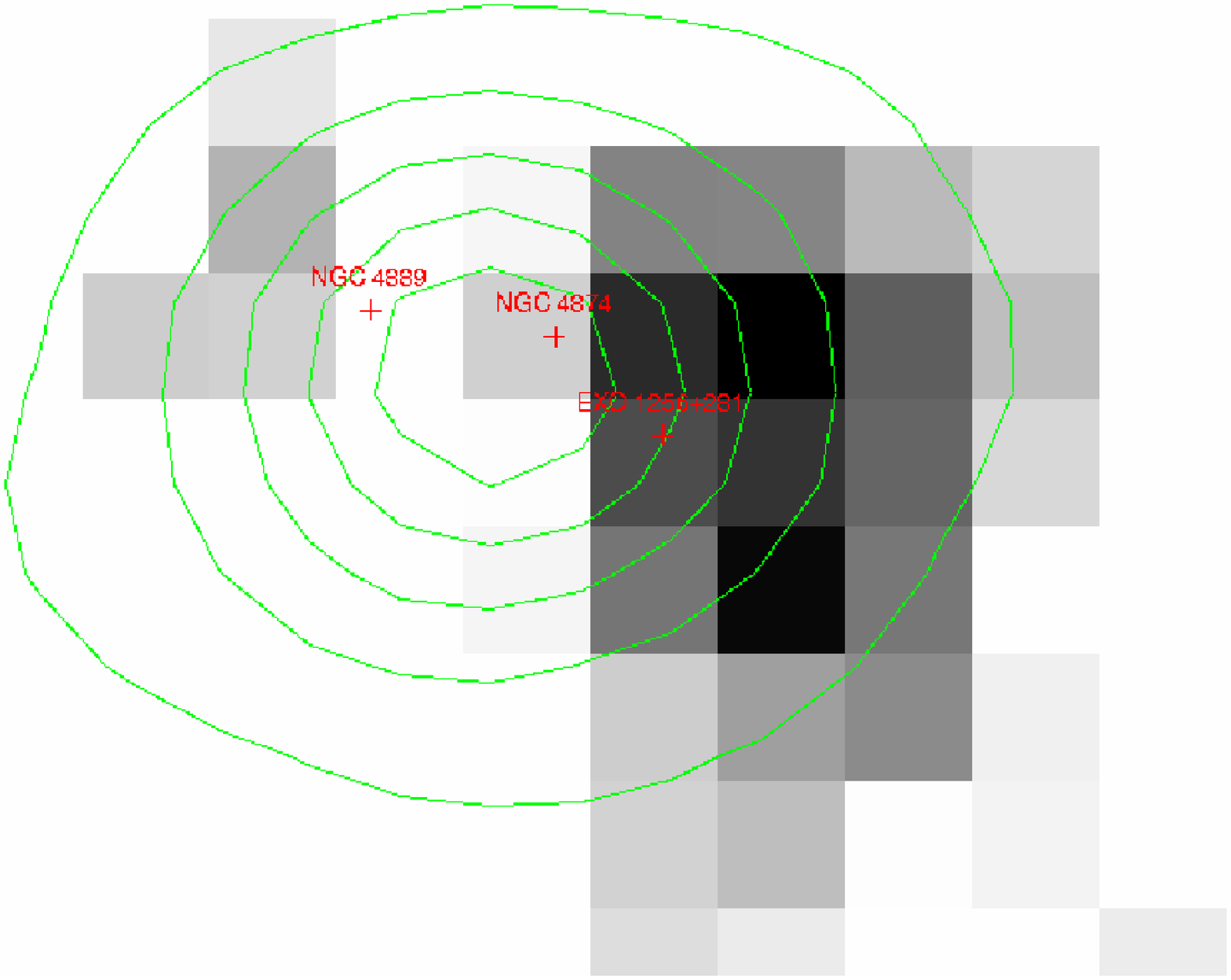}}
     \caption{Left: \intgr\ mosaic image with contours from \xmm\ overlayed. The \xmm\
image is smoothed with a gaussian of the width 12' to match the angular
resolution of \intgr. Right: the residuals after the subtraction of the \xmm\ profile from the \intgr\ image (see text). The South-West excess in the residual image is apparent.}
     \label{substraction}
\end{figure*}

A comparison of the 18-30~keV image with the image in 1-10~keV band obtained with \xmm\
is shown in Fig. \ref{substraction}. The left panel of this figure shows the \intgr\ image with the overlayed contours obtained by smoothing the \xmm\ image with the ISGRI PSF (modelled as a gaussian of full width 12'). The right panel of the figure shows the residuals of the \intgr\ mosaic image after substraction of the smoothed \xmm\ image, renormalized in a way that the difference between \intgr\ and \xmm\ flux cancels at the maximum of the \xmm\ emission. One can clearly see that significant residuals are left in the South-West (SW) part of the \intgr\ source after the substraction. This indicates that the hard X-ray source detected by \intgr\ is more extended in the SW direction than the \xmm\ source.

\subsection{Source morphology from mosaic image}
\label{morph}

\begin{figure}
\resizebox{\hsize}{!}{\hbox{
\includegraphics[width=2.5cm,height=2.5cm]{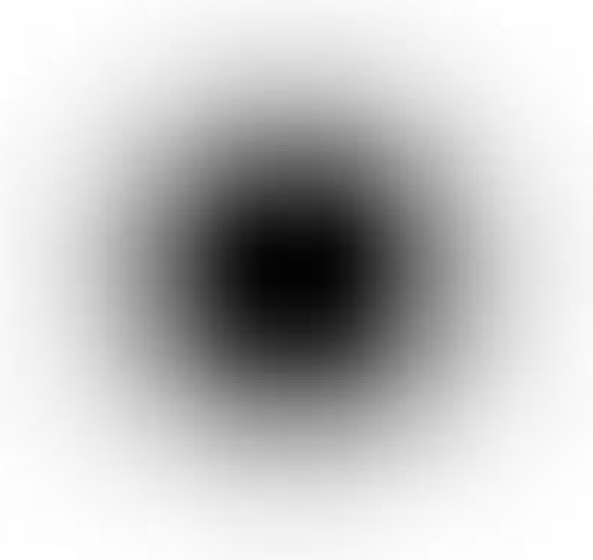}
\includegraphics[width=2.5cm,height=2.5cm]{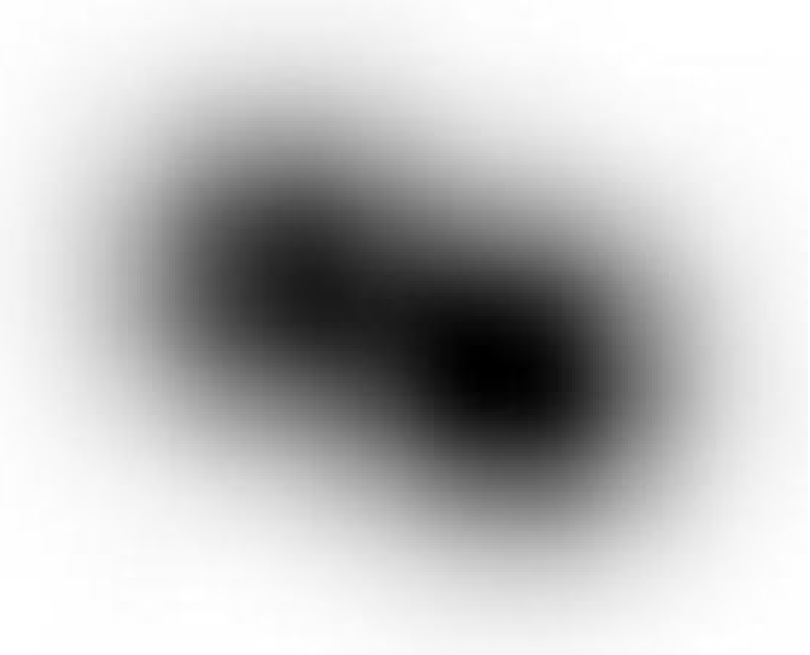}
\includegraphics[width=2.5cm,height=2.5cm]{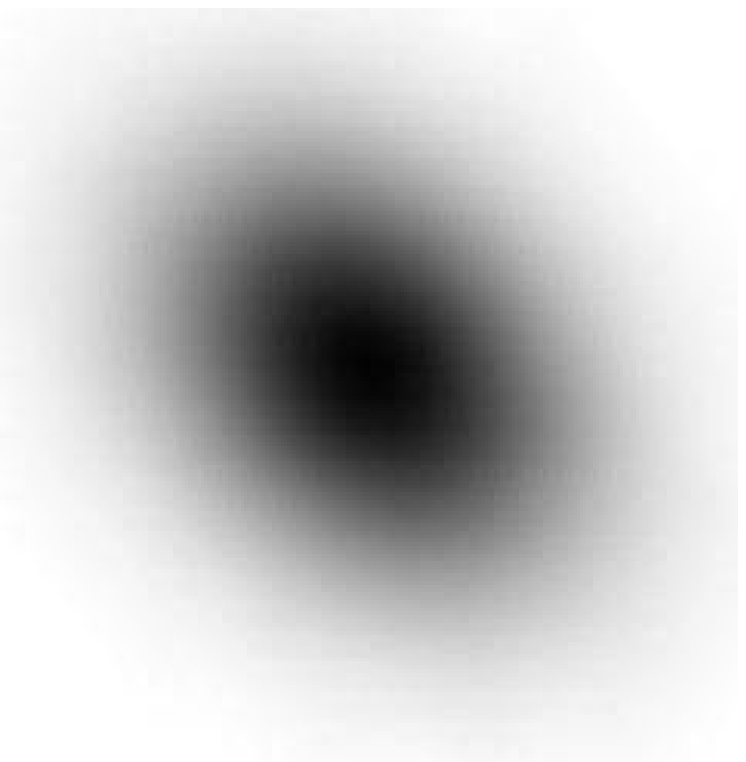}
\includegraphics[width=2.5cm,height=2.5cm]{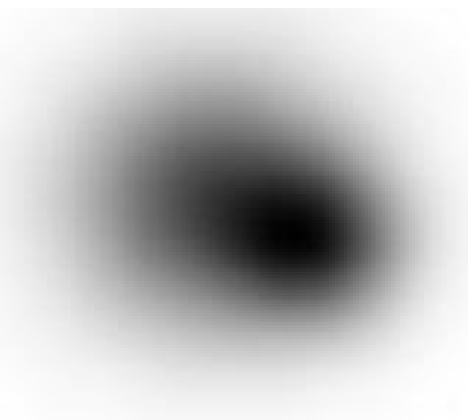}}}
\caption{Comparison between the best fits to \intgr\ mosaic image (see Fig. \ref{figosa}) with a standard point source PSF (left), the PSF of 2 point sources (second from the left), the PSF of an extended source with the shape of an ellipse (third from the left) and the PSF of a source with the morphology of the Coma cluster in the 1-10 keV plus an additional point source (right).}
\label{psf}
\end{figure}

To clarify the nature of the SW extension of the \intgr\ source
we attempted to fit the ISGRI image assuming different surface brightness distributions $I(\vec r)$ of the hard X-ray source. Specifically, we consider the following possibilities:
\begin{itemize}
\item \textbf{Model 1:} A single point source given by a Gaussian with the half-width equal to the size of the PSF of ISGRI,
\begin{equation}
\label{eq:ps}
I(\vec r)=A\exp\left[-\frac{(\vec r-\vec r_0)^2}{2\sigma^2}\right],
\end{equation}
where  $\vec r_0$ is the central vector position of the source which is left free while fitting;
\item \textbf{Model 2:} A superposition of two point sources with overlapping PSFs,
\begin{equation}
I(\vec r)=A\exp\left[-\frac{(\vec r -\vec r_1)^2}{2\sigma^2}\right]+
B\exp\left[-\frac{(\vec r -\vec r_2)^2}{2\sigma^2}\right],
\end{equation}
where $\vec r_1,\vec r_2$ are the positions of the two sources which are left free;
\item \textbf{Model 3:} An ellipse-shaped extended source with the surface brightness profile
\begin{equation}
I(\vec r)=A\exp\left[-\frac{\left((\vec r-\vec r_3)\cdot\vec n\right)^2}{2\sigma_1^2}-\frac{\left((\vec r-\vec r_3)\times\vec n\right)^2}{2\sigma_2^2}\right],
\label{mod3}
\end{equation}
where $\vec n$ is the unit vector in the direction of the major axis of the
ellipse, $\vec r_3$ is the position of the centroid of the ellipse and $\sigma_1,\sigma_2$
are the sizes of the major and minor axes of the ellipse which are all left free while
fitting;
\item \textbf{Model 4:} A superposition of an extended source with morphology of the core of Coma cluster in the 1-10~keV energy band (the surface brightness profile described by a
Gaussian convolved with the ISGRI PSF, i.e. a Gaussian with a half-width of $10'$) and of an additional point source
\begin{equation}
I(\vec r)=A\exp\left[-\frac{(\vec r-\vec r_c)^2}{2\sigma_3^2}\right]+B\exp\left[-\frac{(\vec r-\vec r_4)^2}{2\sigma^2}\right],
\end{equation}
where $\vec r_c$ is the position of the centroid of the soft X-ray emission and $\sigma_3=10'$. $\vec r_4$ is the position of the additional point source which is left free while fitting, and $\sigma$ is fixed to the half-width of the PSF in the same way as in model 1.
\end{itemize}
We fitted the $21\times 21$~pixels ($105'\times 105'$) part of the image around the catalog
position of the Coma cluster minimizing the $\chi^2$ of the fit, defined as
\begin{equation}
\chi^2=\sum_{i,j=1}^{21}\frac{\left({\rm IMG}(i,j)-I(i,j)\right)^2}{{\rm VAR}(i,j)}.
\end{equation}
where IMG$(i,j)$ and VAR$(i,j)$ are the values of intensity and variance in a given
image pixel $(i,j)$. The best fit results for the four models are shown in Fig. \ref{psf}.

Fitting the intensity image with a single point source (Model 1),
we find that the source position (the vector $\vec r_0$ in (\ref{eq:ps})) is
shifted compared to the centroid of the \xmm\ image in the SW direction
by $\Delta$RA$=3.9'$, $\Delta$DEC$=0.5'$. The rather high reduced $\chi^2$ of the fit with a point source model, $\chi_{red}^2=1.91$ (for 437 degrees of freedom) indicates that the single point source model does not provide a good description of the source morphology, confirming the analysis shown in Fig. 2.

Fitting the source morphology with a two point sources (Model 2), one finds a better
reduced $\chi^2_{red}=1.33$. The two point sources model provides the possibility of finding the
direction of extension of the source. Namely, the best fit is provided by the model
in which the two point sources are situated at RA$_1=195.04 \pm 0.01$, DEC$_1=28.00  \pm 0.01$ and RA$_2=194.77\pm0.01$, DEC$_2=27.90\pm 0.01$. The angular distance between the two point sources is $d_{12}=\left|\vec r_1-\vec r_2\right|=15.3'\pm 0.6'$. The ratio of intensities of the two sources is $A/B=0.84$. The distance between the two point sources is larger than the size of the PSF of ISGRI, which confirms again that the source cannot be described by Model 1.

The best fit to the morphology of the source is found when fitting the image
with the model of an elliptically-shaped source (Model 3). The fit results in a reduced
$\chi^2_{red}=1.23$. The parameters of the best fit model are $\sigma_1=16.8'\pm 0.5'$,
$\sigma_2=11.7'\pm 0.4'$ and the coordinates of the centroid of the ellipse, $\vec r_3$, RA$=194.89\pm 0.01$, DEC$=27.94\pm0.01$. The direction of the major axis of the ellipse, $\vec n$, is inclined at the angle $\theta=61\pm 4^\circ$. One can see that the fitted position of the centroid of the ellipse is shifted in the same direciton as the position of the single point source fit, $\vec r_0$. The size of the major axis of the ellipse, $\sigma_1$ is roughly equal to $d_{12}$.

\begin{figure}
\resizebox{\hsize}{!}{\includegraphics{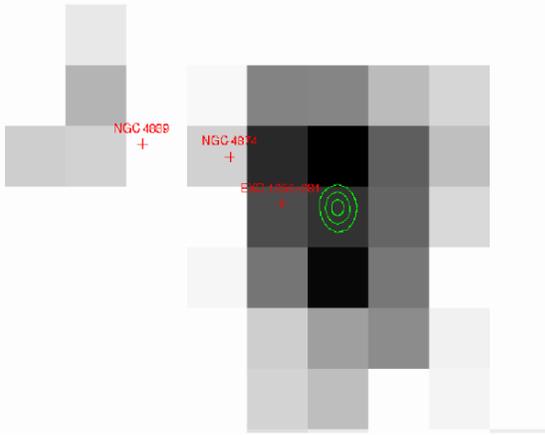}}
\caption{Residual image after substraction of the scaled \xmm\ image from the \intgr\ mosaic image. The position of the fitted point source from Model 4 is displayed with 1$\sigma$, 2$\sigma$ and 3$\sigma$ error contours. The positions of the brightest AGNs emebedded in the cluster is also displayed.}
\label{difference}
\end{figure}

The fit by an extended source of the shape of the Coma cluster in the 1-10 keV band plus an additional point source (Model 4) gives a good fit, $\chi^2_{red}=1.36$. The additional point source is found at RA=$194.71\pm0.01$ and DEC=$27.87\pm0.01$, which is located 6.1' away from the quasar EXO 1256+281. Fig. \ref{difference} shows the position of the fitted point source on the residual image with 1$\sigma$, 2$\sigma$ and 3$\sigma$ error contours.

\section{A method to analyse extended sources with a coded mask instrument}
\subsection{Description of the method}
\label{secpif}

The imaging capabilities of ISGRI make it possible to obtain information on the morphology of the hard X-ray emission of the Coma cluster and other slightly extended sources for the first time. However, the standard Offline Scientific Analysis (OSA) software distributed by ISDC \citep{cour} is optimized for point sources, and is not well suited for slightly extended sources. We present here a method based on Pixel Illumination Fraction (PIF) which extracts the properties of slightly extended sources
with a coded mask instrument \citep[see also][]{ren1}.

For a coded mask instrument, the sky images and, in particular the mosaic image studied
in the previous section, are produced by the backprojection of the shadow patterns
cast by the sources in the field of view on the plane of the sky. The shadow
pattern produced when observing a FOV containing $n$ sources is a superposition of the shadow patterns of all the individual sources,
\begin{equation}S(x,y)=\sum_{i=1}^n f_i\cdot\mbox{PIF}_i(x,y)+B(x,y),\label{pif}\end{equation}
where $f_i$ and PIF$_i(x,y)$ are the flux, respectively the shadow pattern (called ``Pixel Illumination Fraction'') of the $i$th source and $B(x,y)$ is the background in the pixel with coordinates $(x,y)$. The Pixel Illumination Fraction gives the fraction of each pixel of the detector illuminated by the source. For a pixel that is completely in the shadow of the mask, the PIF will be 0, whereas in the case of a fully illuminated pixel, the PIF will be equal to 1. It is understandable that the PIF of an extended source is different from that of a point source, since some pixels might be illuminated by only a fraction of the total extension of the source, which cannot happen for a point source. Thus to describe an extended source properly, one has to create an appropriate model for it. Our method to create such a model is the following: we create a grid of positions on the sky covering the extended source, compute the PIF for all the positions of this grid, and then average the PIFs, weighted
by a model surface brightness for the source, e.g. a spherical isothermal $\beta$-profile,
\begin{equation}I(\vec r)\propto\frac{1}{\left(1+\frac{{\vec r}^2}{a^2}\right)^\beta}.\label{profile}\end{equation}

\begin{figure}[hbt]
\resizebox{\hsize}{!}{\hbox{
\includegraphics[width=5cm]{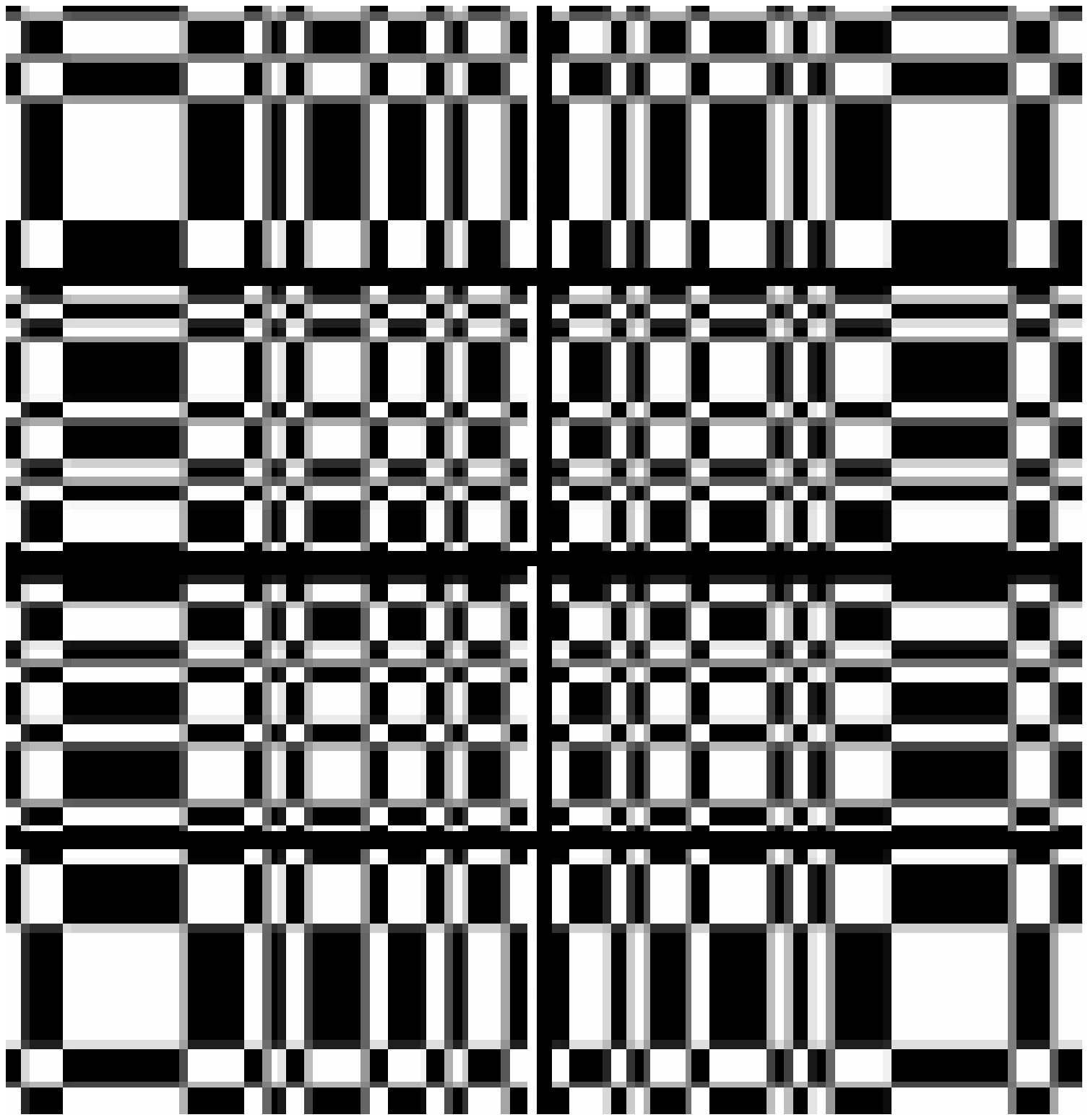}
\includegraphics[width=5cm]{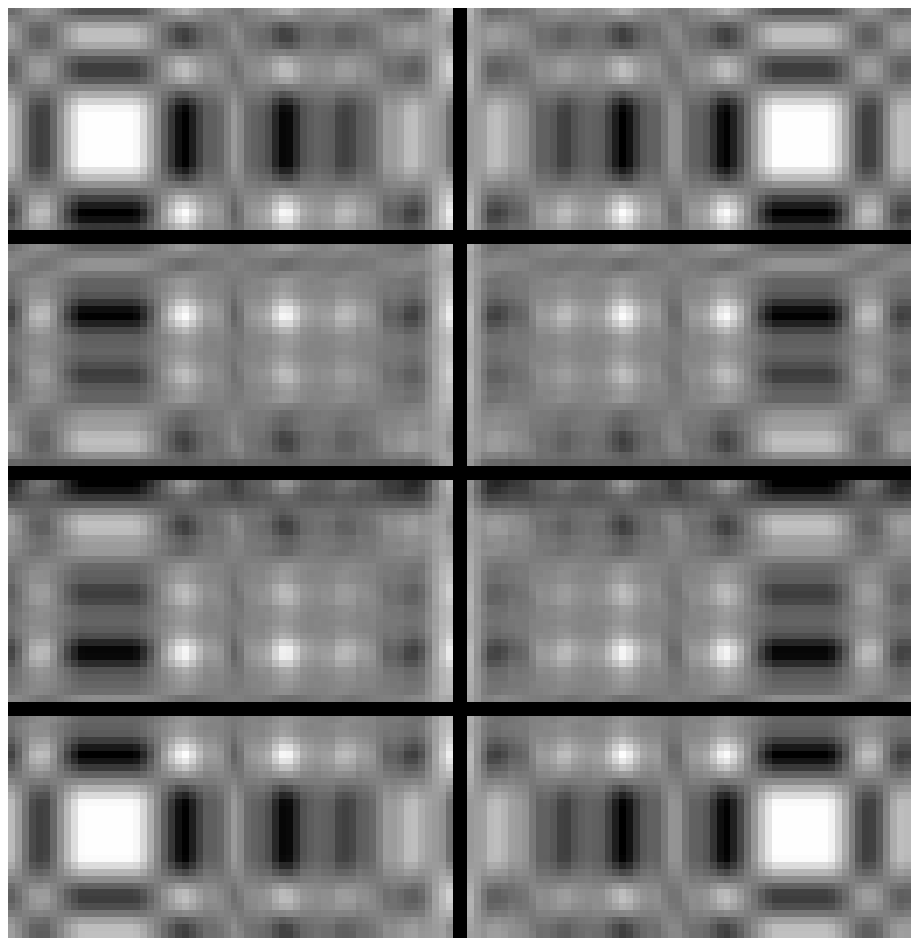}}}
\caption{The Pixel Illumination Fraction (PIF) of the IBIS/ISGRI instrument for a source in the center of the field of view (white=1, black=0): \emph{left:} for a point source; \emph{right:} for an extended source with a surface brightness described by an isothermal $\beta$-profile, with a source size $a=30$ arcmin.}
\label{figpif}
\end{figure}

Fig. \ref{figpif} shows the PIF for an on-axis source, in the case of a point source and for an extended source described by the model of Eq. \ref{profile} with $a=30'$.

Since the standard OSA tools always use the PIF for a point source, in the case of an extended source they do not extract fluxes and spectra correctly. To extract the flux of an extended source, we created a tool that fits the detector image (or \emph{shadowgram}) with any kind of PIF, either point-like or extended with a given surface brightness model. Our tool creates a correct PIF for every source in the FOV, and then fits the shadowgram to the model described in Eq. (\ref{pif}), where 1 or more sources might be extended. To check the coherence of our method, we have simulated the shadowgram one can expect with 2 sources, one extended in the middle of the FOV with a flux $F_1=40$ counts per fully illuminated pixel (cpp hereafter) and one point-like in the corner of the FOV with a flux $F_2=80$ cpp, with a gaussian background, and extracted the fluxes of the 2 sources with our tool. The results gave us $F_1=40.0\pm0.7$ cpp and $F_2=80.4\pm1.2$ cpp, which shows that our method is indeed extracting fluxes properly.

This method can also be used to extract spectra: we analyzed the data with the standard OSA tools to create shadowgrams in all the energy bands desired for the spectrum, and then used our fitting tool to extract a flux with the correct PIF in all the energy bands, and reconstruct a spectrum (see Sect. \ref{secspec}).

It is important to note that the use of a PIF-based method for weak extended sources could be complicated because of the specifics of operation of the \intgr\ satellite. Namely, each \intgr\ observation is split into several kilosecond-long intervals of continuous data taking, called Science Windows (ScWs). The statistics of the signal from a weak source in each ScW is below the background statistics. Moreover, the statistical properties of the low-statistic signal in the ISGRI imager are not well known. This means that the use of gaussian statistics for the fitting of the shadowgrams is not completely correct, and can give wrong results. On the contrary, the method of analysis based on the mosaic image is better suited for the analysis of weak extended sources. In any case, the two methods should be considered complementary, and for this reason we present results based on both methods in this paper.

\subsection{Hypothesis of multiple point sources}
\label{secpoint}

One possible explanation for the shape of the mosaic image (Fig. \ref{figosa}) is that one or several point sources contribute to the observed flux. We analyse this possibility here. Candidate sources are the cluster itself and bright AGNs at close angular distances. The three brightest AGNs in soft X-rays are the central radio galaxy NGC 4874, NGC 4889 and the QSO EXO 1256+281. Fig. \ref{figosa} shows the position of these 3 point sources on the ISGRI image.

We used the method described in Sect. \ref{secpif} to fit the shadowgram in the 18-23 keV band by the PIF of 3 point sources at the position of the sources given above. We extracted the flux of each of these 3 sources, and created a model of the deconvolved image with the flux found. 

\begin{figure}
\resizebox{\hsize}{!}{\includegraphics{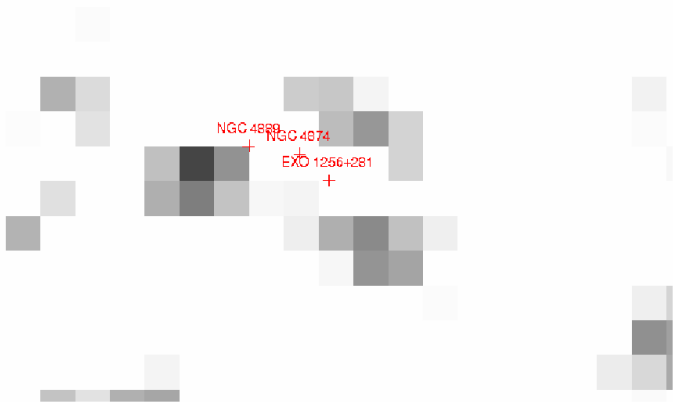}}
\caption{Substraction between the mosaic image in the 18-23 keV band and the model obtained with the fit of the data with the sum of NGC 4874, EXO 1256+281 and NGC 4889. While the center of the emission is well described, the outer parts of the image show a deviation of up to 5.5$\sigma$ from the model.}
\label{figmod}
\end{figure}

Fig. \ref{figmod} shows the result of the substraction of the fitted model from the mosaic image in the 18-23 keV. We can see on the image that the outer parts of the source are not well described by the model, specially the region at the bottom right and the left, where 5.5$\sigma$ excesses are observed in the mosaic compared to the model image. This analysis suggests that the emission seen by \emph{INTEGRAL} is more extended than what can be explained by the superposition of the brightest AGNs in
the cluster.

\begin{figure}
\resizebox{\hsize}{!}{\includegraphics{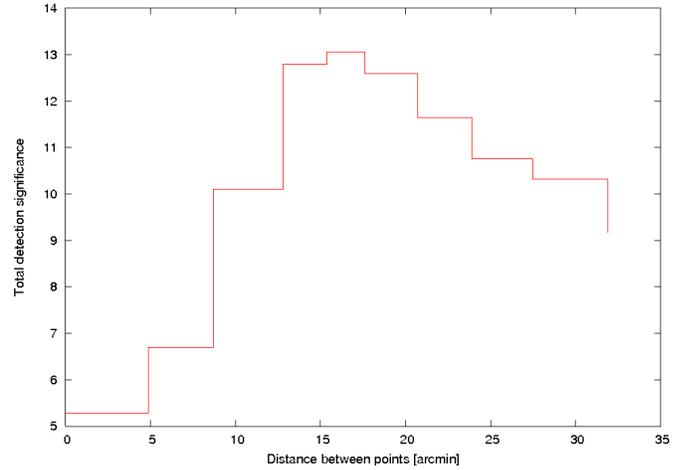}}
\caption{Computation of the total detection significance in the 18-23 keV band as a function of the distance between 2 fake sources along the major axis of the ellipse.}
\label{detsig_2sources}
\end{figure}

To confirm this, we fit the data with 2 fake point sources placed along the major axis of the ellipse, and computed the detection significance for different distances of the 2 fake sources. In the case of a single point source, we would expect the detection significance to drop when we increase the distance between the 2 fake sources, whereas in the case of 2 sources whose PSF overlap, we expect the detection significance to peak at the distance between those 2 sources. The
position of the 2 sources for which the detection significance peaks will therefore allow us to compare this result with the position of known point sources inside the cluster. If there is a possible point source counterpart in soft X-rays, this will give an indication that contamination of the hard X-ray flux by point sources is likely. The result of this computation is shown in Fig. \ref{detsig_2sources}.

We can see on Fig. \ref{detsig_2sources} that the detection significance peaks at a distance of $17 \pm 2$ [arcmin], which is fully compatible with the result obtained from the mosaic in Sect. \ref{secmos}. Since we expect the thermal emission from the cluster to dominate in this energy range, we would expect the two points to be located on the centre of the cluster (that lies 1' south of NGC 4874) and one of the other AGNs. However, the distance between the centre of the cluster and EXO 1256+281
(4.6') or NGC 4889 (8.9') is at least a factor of 2 smaller than the distance we found. This is a strong argument in favour of extended emission.

\subsection{Extended emission analysis}

\begin{figure}
\resizebox{\hsize}{!}{\includegraphics{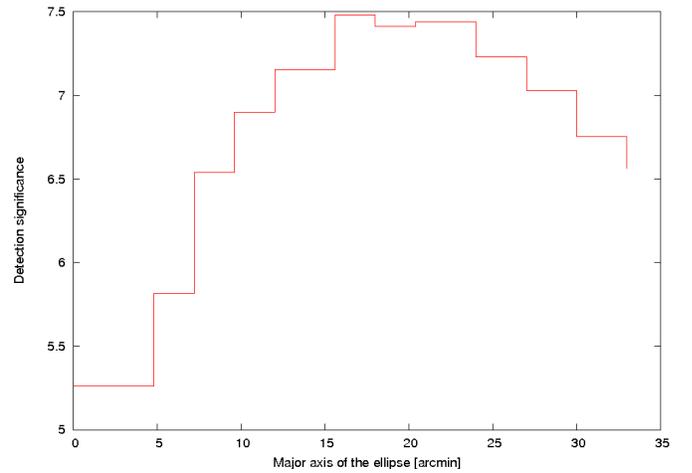}}
\caption{Detection significance of the source computed with an ellipse-shaped PIF in the 18-23 keV band as a function of the major axis of the ellipse.}
\label{detsig_ellipse}
\end{figure}

We used the same method to analyse the ISGRI data with the assumption that the emission is indeed diffuse emission. We used the angle of the ellipse found from the mosaic analysis (see Sect. \ref{secmos}) to create PIFs of an extended source with the method explained in Sect. \ref{secpif}, weighted by a surface brightness given by Model 3 (see Eq. (\ref{mod3})). We used our tool to fit all the data with this model for several different values of $\sigma_1$ and $\sigma_2$ with a ratio $\frac{\sigma_1}{\sigma_2}=1.43$ fixed by the results of the image fitting, and finally computed the detection significance of the source for all the different models. The result is shown in Fig. \ref{detsig_ellipse}.

We can see in this Fig. that the detection significance also peaks at a major axis size of about 17 [arcmin], which is again comparable to the results obtained from the mosaic. We can see that the maximum detection significance is lower than in the case of 2 point sources, but this does not mean that the model of 2 point sources gives a better representation of the data: indeed, the error n the fit increases with the size of the source, because the contours of the smallest holes of the mask become more
and more unclear (see Fig. \ref{figpif}), and thus the imaging method is less accurate. It is thus irrelevant to compare directly Figs. \ref{detsig_2sources} and \ref{detsig_ellipse}, but both show independantly the fact that the source is not point-like, and provide a measurement of the apparent size of the source in hard X-rays.

From this analysis, we conclude that if we use a source size smaller than the limit of 17', we lose a part of the total flux of the cluster, and thus the detection significance increases with the size. In the opposite case, we collect more background, so the detection significance starts to drop. The best estimate of the total flux of the cluster is therefore given at a major axis size of 17 [arcmin], which gives $F_{tot}=0.31\pm0.04$ counts/s in the 18-30~keV band. From now on, we will use this model to
extract fluxes and spectra.

We also performed another complementary analysis: we fitted the data with 4 sources, i.e. the extended model described above and the 3 AGNs. The fit converges to a solution that puts 80\% of the flux in the extended emission. This model cannot be used to extract a flux, because the fluxes of the 4 sources become strongly anti-correlated. However, it means that the extended source model describes the data better than the sum of point sources.

\section{Spectral analysis}
\label{secspec}

\begin{figure}
\resizebox{\hsize}{!}{\includegraphics[angle=270]{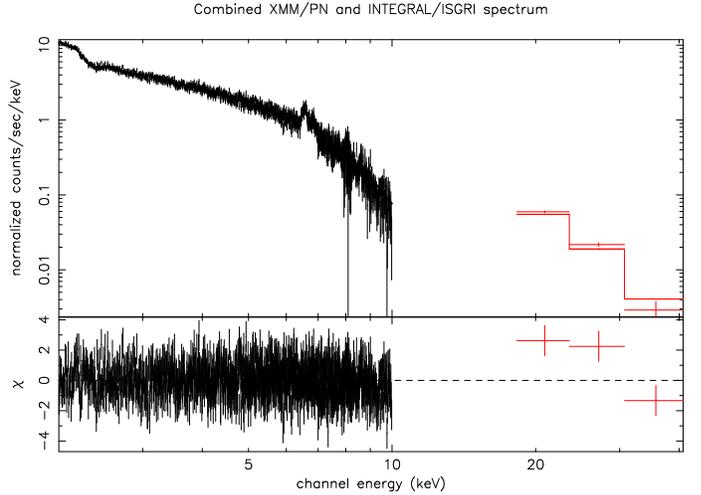}}
\caption{Combined \emph{XMM}/PN and \emph{INTEGRAL}/ISGRI spectrum. The PN spectrum is fitted by a MEKAL model at a temperature $kT=7.9\pm0.1$ keV. The bottom plot shows the residuals of the best fit compared to the data. There is no obvious deviation to this spectrum from the ISGRI data points up to 40 keV.}
  \label{spectrum}
\end{figure}

To extract the spectrum of the source correctly, we used the method described in Sect. \ref{secpif}. We used the hard X-ray shape of the source extracted from the \emph{INTEGRAL} image to create a PIF covering the whole size of the source, and extracted the flux from the shadowgram of each pointing in 3 different energy bands: 18-23, 23-30 and 30-40 keV (the source is not detected at higher energies). We then performed a weighted sum over the flux extracted from all pointings to get a total spectrum in these 3 energy bands. We also extracted the \emph{XMM}/PN spectrum of the cluster with the background substraction method from the Birmingham group, in a region chosen such that the \xmm/\intgr\ intercalibration factor is equal to 1. Finally, we fitted this spectrum in XSPEC with the MEKAL model \citep{kaastra}. We extrapolated the fitted model to higher energies and compared it to the data points obtained with ISGRI. The result is shown in Fig. \ref{spectrum}.

Because of the very low statistics at energies above 30 keV, we are not able to confirm or deny the presence of a non-thermal hard X-ray excess emission. Indeed, the extended nature of the source makes it difficult to extract a significant spectrum up to high energies, since the already low statistics is spread over several sky pixels. A longer exposure time is therefore required to make conclusions on the presence or not of a hard X-ray excess emission above 30 keV.

Although there is strong evidence that EXO 1256+281 cannot explain the \intgr\ SW extension, the angular distance between the fitted position and EXO 1256+281 (6.1') is close to the half-width of the ISGRI PSF. Hence, we further investigate this point by extracting the spectral properties of this object, assuming the identification of the additional point source with EXO 1256+281.

\begin{figure}
\resizebox{\hsize}{!}{\includegraphics[angle=270]{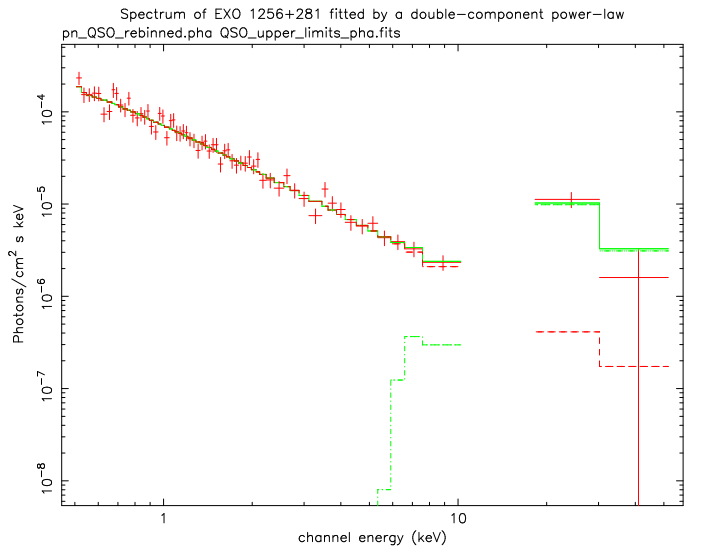}}
\caption{\xmm\ spectrum of the quasar EXO 1256+281 with one ISGRI data point obtained by fitting the south-west excess in hard X-rays by a point source at the position of EXO 1256+281. The ISGRI upper limit in the 30-50 keV band is shown. The spectrum is fitted by the sum of a power law (dashed red) and a heavily absorbed component dominating the flux at higher energies (dashed green).}
\label{exospec}
\end{figure}

To this end, we fitted the excess in the ISGRI image by a point source at the position of EXO 1256+281, and extracted the flux for this source. Fig. \ref{exospec} shows the \xmm\ spectrum of EXO 1256+281 with the ISGRI point extracted using this method. The \xmm\ spectrum is well fitted by a simple powerlaw with the photon index $\Gamma=1.63\pm0.07$ and 0.5-10~keV flux $5\times10^{-13}\mbox{ ergs }\mbox{cm}^{-2}\mbox{ s}^{-1}$. The flux of the source in the \intgr\  energy band is much higher than the extrapolation of the powerlaw found in \xmm\ data. This implies the presence of an additional spectral component which dominates the source above 10~keV. To model this component, we added a heavily absorbed power law typical of Seyfert 2 galaxies to the fit. Taking into account the upper limit in the 30-50 keV band, the fit gives $n_H=(4.0\pm1.7)\times10^{24}\mbox{ cm}^{-2}$, and $\Gamma\geq 3.0$ to match the upper limit. This index is much steeper than the unified Seyfert 2 spectral index ($\Gamma=1.79$ with a dispersion $\sigma=0.23$, \citet{risaliti}), and hence, the properties of this source would be very unusual for a Seyfert 2 galaxy. From this statement together with the imaging arguments presented in the previous sections, we conclude that the contribution of known point sources to the observed flux is very unlikely between 10 and 40 keV.

Assuming that the SW excess is due to the presence of diffuse emission, which gives the best representation of the data, one can try to constrain the properties of the gas needed to explain the emission from this region.   We attempted to make a joint fit of the 1-10~keV spectrum extracted from a circle of a radius of 6' centered at the position of the SW excess and the 18-50~keV spectrum extracted from the SW region from the \intgr\ mosaic image in the same way as in the case of an additional point source. Assuming that all the flux from the SW region comes from a higher temperature plasma, we fitted the data with the thermal bremsstrahlung model. This results in a temperature of $kT=12\pm2$ keV. The corresponding estimate of the emission measure (EM) is $0.16\leq\mbox{ EM }\leq0.26\mbox{ cm}^{-6}\mbox{ pc}$, which is reasonable for external regions of the cluster. We can thus conclude that the presence of a hotter region (10 keV$<kT\leq14$ keV) can explain the extension found in the ISGRI mosaic image. 

\section{Discussion}

In this paper, we have used the IBIS/ISGRI instrument on board the \emph{INTEGRAL} satellite to investigate the hard X-ray emission from the Coma cluster. We presented a method based on Pixel Illumination Fraction (PIF) to analyse extended sources with a coded mask instrument (section \ref{secpif}), and we have shown that the Coma cluster indeed appears like an extended source for ISGRI. Assuming that the emission seen by \emph{INTEGRAL} is extended, we have compared the ISGRI mosaic image with the soft X-ray image from \emph{XMM-Newton}, and shown that there is a displacement between them: the \emph{INTEGRAL} image is displaced towards the south-west, i.e. in the direction of the NGC 4839 group, that is currently merging with the main cluster.

The origin of the extended emission from this region is not clear, but we have investigated two possible explanations for the excess in the image: an additional heavily absorbed point source embedded in the cluster, and an extended region where large-scale shocks occur.

The first model for the hard X-ray excess in the South-West region of the cluster is the presence of a highly absorbed additional point source appearing at higher energies. We have used the imaging capabilities of the instrument to investigate this possibility: we have shown that the shape of the residual image after substraction of the \xmm\ surface brightness profile from the ISGRI image does not coincide with any known X-ray point source, and that the only possible candidate EXO 1256+281 is located more than 6 arcmin away from the best position found in the \intgr\ image, which makes it an unlikely counterpart. We have also extracted the soft X-ray spectrum of this source and the flux of the south-west region in \intgr\ data, and shown that this spectrum is not compatible with a highly absorbed Seyfert II galaxy. Indeed, the source is not detected in the 30-50 keV band, which is the most sensitive energy band of ISGRI. This implies a spectral index $\Gamma\geq3.0$, which is too steep for a highly absorbed Seyfert 2 galaxy. As a conclusion, we claim that contribution of a very hard point source embedded in the cluster to the observed spectrum is highly unlikely.

We cannot exclude the possibility that the South-West excess in hard X-rays is due to one or few unknown sources that would emit predominantly in this energy band. If this is the case, these sources must be highly absorbed ($n_H\geq4\times10^{24}\mbox{ cm}^2$), and have a steep spectral index $\Gamma\geq3.0$. We note that these characteristics are unlikely, because most of the highly absorbed sources discovered by \intgr\ show a much harder spectrum.

In the scenario of a merging event between the Coma cluster and the NGC 4839 group, we expect a shock front to be created in the region where the gas of the two clusters collides. Our imaging analysis shows that the hard X-ray emission seen by \emph{INTEGRAL} is extended in the direction of the NGC 4839 group, which is a good indication that the emission we see is indeed coming from a region where large-scale shocks occur. If this explanation is correct, we expect the plasma in this region to be hotter, and hence to have a harder spectrum in X-rays. The temperature map of the cluster \citep{neumann} shows a region that has low surface brightness in X-rays, but is the hottest region of the cluster ($kT\geq 10$ keV). This region coincides with the position of the hard X-ray South-West extension found by \intgr, and we can thus associate the hard X-ray excess in this region discussed in Sect. \ref{secspec} with emission from a very hot region of the cluster ($kT\leq14$ keV). This result is consistent with the temperature found in the merging region of the distant cluster Cl J0152.7-1357 \citep{maughan}, which shows that such a high temperature is possible and might indeed be the signature of a merger. Assuming that the hot region is roughly spherically symetric and has an angular size of $\sim 6'$ at a distance of 100 Mpc, we deduce that the density of the hot gas is $n_{hot}\sim 10^{-3}\mbox{ cm}^{-3}$. Given that the South-West excess in the ISGRI image appears to be extended, \intgr\ data appear to confirm this scenario.

\label{secdisc}
\begin{acknowledgements}
We would like to thank M. Chernyakova for help with \xmm\ data analysis and helpful comments. This work is based on observations with INTEGRAL, an ESA project with instruments and science data centre funded by ESA member states (especially the PI countries: Denmark, France, Germany, Italy, Switzerland, Spain), Czech Republic and Poland, and with the participation of Russia and the USA.
\end{acknowledgements}

\bibliographystyle{aa} 
\bibliography{077268} 
\end{document}